\documentclass[12pt,prd,aps,amssymb,amsmath,tightenlines,showpacs,a4paper,nofootinbib]{revtex4}
\def\be{\begin{eqnarray}}
\def\ee{\end{eqnarray}}
\def\bea{\begin{eqnarray}}
\def\eea{\end{eqnarray}}
\def\beax{\begin{eqnarray*}}
\def\eeax{\end{eqnarray*}}
\def\half{\frac{1}{2}}

\def\ve{\varepsilon}

\usepackage{color}
\usepackage{graphicx}

\begin{document}
\title{An angle on invisibility}

\author{H.~F.~Jones$^\dag$ and M.~Kulishov$^\ddag$}
\affiliation{
$\phantom{.}^\dag$Physics Department, Imperial College, London SW7 2BZ, UK\\
$\phantom{.}^\ddag$HTA Photomask, 1605 Remuda Lane, San Jose, CA 95112, USA}
\date{\today}

\begin{abstract}
The $PT$-symmetric optical grating with index profile $e^{i\beta z}$ has been shown to have the interesting
property of being essentially invisible for light incident from one side, while possessing greatly enhanced
reflection at a particular wavelength for light incident from the other side. We extend a previous analysis
of this grating to obtain an analytic solution for the case when the grating is embedded on a substrate, with
different refractive indices on either side. We also generalize the previous case of normal incidence to incidence
at an arbitrary angle. In that case the enhanced reflection occurs at a particular angle of incidence for a given wavelength.
Finally we discuss how the grating may be used to give lasing.
\end{abstract}


\maketitle
\section{Introduction}

The ideas of $PT$ symmetry, originally introduced in the context of quantum mechanics\cite{BB}-\cite{AMh}, have recently
led to rapid developments in the apparently unconnected field of classical optics\cite{op1}-\cite{op9}. The connection
arises from the fact that when one makes the paraxial approximation for the equation of propagation of an electromagnetic wave the resulting equation is formally identical to the Schr\"odinger equation, but with different interpretations for the symbols appearing therein. In particular, the role of time in the Schr\"odinger equation is taken by the longitudinal coordinate $z$, while that of the quantum-mechanical potential is taken by variations of the refractive index of the medium in the transverse $x$ direction.  $PT$-symmetry deals with potentials that are not Hermitian, which translates in optics to complex refractive indices. It is extremely common for the refractive index to have a positive imaginary part, corresponding to loss, but a negative imaginary part can also be implemented by optical pumping, leading to gain. $PT$-symmetry requires that loss and gain be balanced in a particular way, namely that $n^*(-x)=n(x)$, or equivalently that ${\rm Re}(x)$ be an even function and ${\rm Im}(x)$ an odd function of $x$. When the $PT$ symmetry is unbroken this leads to real propagation constants, i.e. no exponential growth or decay even in the presence of gain and loss.

Although this connection was first made explicit in Ref.~\cite{op1}, the exotic properties of materials with combined gain and loss
were previously explored in Refs.~\cite{Poladian} and \cite{Berry}. In those papers the index modulation was in the $z$ direction rather than the transverse $x$ direction. In that case the equation of propagation for a given component of the electric field is just the scalar Helmholtz equation
\be\label{Helmholtz}
\left[d^2/dz^2 + k^2(n(z)/n_0)^2\right] E(z)=0,
\ee
which can be compared to the time-independent Schr\"odinger equation. This situation was highlighted in the paper by Lin et al.\cite{Lin} (see also Ref.~\cite{MKK}). For the particular case when the variation in $n(z)$ is a pure complex exponential proportional to $e^{i\beta z}$ the metamaterial exhibits to a very good approximation the phenomenon of ``unidirectional invisibility" for normal incidence, with perfect transmission and zero reflection from the left. On the other hand, for right incidence, the concomitant property is a greatly enhanced reflectivity, sharply peaked at $k=\beta$.

In Refs.~\cite{Longhi} and \cite{HFJ1} an analytic solution was found for the scattering coefficients in terms of Bessel functions, showing
the limitations of the coupled-mode approximation used in Ref.~\cite{Lin}. It was shown that for the parameters used in that paper there were small deviations from invisibility, and that the property broke down completely for much longer lengths of the grating.

These papers were concerned with the one-dimensional situation described by Eq.~(\ref{Helmholtz}), that is, for normal incidence on the lattice (as shown in Figs.~1 and 2 below). However, it is of interest to generalize this situation in two ways. First one can consider incidence at an angle, and secondly the situation when the grating is superimposed on a material of different refractive index than those on either side. This problem has recently been addressed in Ref.~\cite{RG}, but in the context of the Bragg approximation method, with analytic expressions obtained for the first three Bragg orders. In the present paper we demonstrate that the method of Ref.~\cite{HFJ1}
can be extended to deal with both these generalizations, yielding analytic expressions for the reflection and transmission coefficients
in terms of modified Bessel functions. Our results can be used to check the approximations used in Ref.~\cite{RG}, but are applicable
for a much wider range of parameters.

In the following Sec.~\ref{Recap} we briefly review the methodology and results of Ref.~\cite{HFJ1}. Then in Sec.~\ref{New}
we generalize the analysis to include non-normal incidence and different refractive indices on either side of the grating. These results are then used in Sec.~\ref{Numerical} to produce graphs of transmission and reflection coefficients, as a function of angle, in a variety of
different configurations. In Sec.~\ref{Mirror} we revert to the one-dimensional situation and consider placing a mirror at one end of the
grating, showing that, because of the enhanced reflectivity of the grating, the cavity can lase when its strength exceeds a certain critical value. Finally, in Sec.~\ref{Conclusions}, we give our conclusions.

\section{Analytic Solution for Restricted Case}\label{Recap}
The set-up dealt with in Refs.~\cite{Lin}, \cite{Longhi}, \cite{HFJ1} is shown in Fig.~1 (for left incidence). We recall here the results of Ref.~\cite{HFJ1} for completeness and comparison with the results in the more general case.
\begin{center}
\begin{figure}[h]
\resizebox{!}{1.5in}{\includegraphics{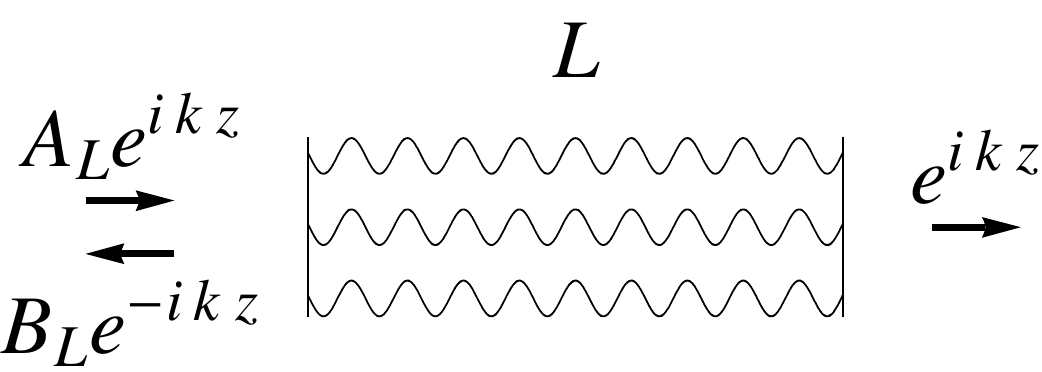}}
\caption{Set-up for propagation from the left}
\end{figure}
\end{center}
The situation is essentially one-dimensional, and Eq.~(\ref{Helmholtz}) reduces to
\be
\frac{d^2E}{dz^2}+k^2(1+2v(z))E=0,
\ee
where $v(z)$ has the special form $v(z)=\half\alpha^2 e^{2i\beta z}$.

Changing variables to $y=(k\alpha/\beta)e^{i\beta z}$, the equation becomes
\be
y^2\frac{d^2E}{dy^2}+y\frac{dE}{dy}-(y^2+k^2/\beta^2)E=0,
\ee
which is the modified Bessel equation, with solution $E=CI_\nu(y)+DK_\nu(y)$,
where $\nu=k/\beta$. In the language of quantum mechanics ($\psi\equiv E$),
\be
\psi(z)=CI_\nu(y)+DK_\nu(y)
\ee
This has to be matched on to $\psi=A_\pm e^{ikz}+B_\pm e^{-ikz}$ at $z=\pm L/2$.

\subsection{Left Incidence}

By requiring that $\psi(L/2)=e^{ikL/2}$ and $\psi'(L/2)=ike^{ikL/2}$ we find that
\beax
C&=&y_+K_{\nu+1}(y_+)e^{ikL/2}\cr
D&=&y_+I_{\nu+1}(y_+)e^{ikL/2},
\eeax
where $y_\pm\equiv \nu \alpha\ e^{\pm i\beta L/2}$,
so that
\be
\psi(z)=y_+\left[K_{\nu+1}(y_+)I_\nu(y)+I_{\nu+1}(y_+)K_\nu(y)\right]e^{ikL/2}
\ee
Both initial conditions are satisfied by virtue of the various recursion relations among the modified Bessel functions and the Wronskian identity\cite{AS}
\be\label{Wronskian}
K_{\nu+1}(y)I_\nu(y)+I_{\nu+1}(y)K_\nu(y)=1/y\ .
\ee

At $z=-L/2$ we have to match $\psi$ with $A_L e^{-ikL/2} +B_L e^{ikL/2}$.
The general formulas are
\beax
A_L e^{ikz}&=& \half[\psi(z)-(i/k)\psi'(z)]\cr
&&\cr
B_L e^{-ikz}&=& \half[\psi(z)+(i/k)\psi'(z)]
\eeax
which give, after some algebra,
\bea\label{LT}
A_L&=& (\half \alpha^2 \nu) e^{ikL}[K_{\nu+1}(y_+)I_{\nu-1}(y_-)-I_{\nu+1}(y_+)K_{\nu-1}(y_-)]\nonumber\\
&&\\
B_L&=& (\half \alpha^2 \nu) e^{-ikL}[-K_{\nu+1}(y_+)I_{\nu+1}(y_-)+I_{\nu+1}(y_+)K_{\nu+1}(y_-)]\nonumber
\eea


\subsection{Right Incidence}
The set-up is shown in Fig.~2, with the transmitted amplitude again normalized to 1.
\begin{center}
\begin{figure}[h!]
\resizebox{!}{1.5in}{\includegraphics{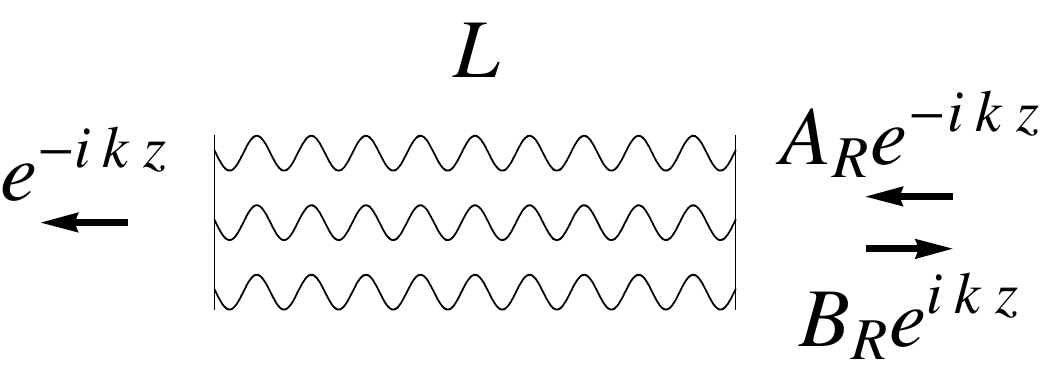}}
\caption{Set-up for propagation from the right}
\end{figure}
\end{center}
The initial conditions are now $\psi(-L/2)=e^{ikL/2}$ and $\psi'(-L/2)=-ike^{ikL/2}$, which give
\beax
C&=&y_-K_{\nu-1}(y_-)e^{ikL/2}\cr
D&=&y_-I_{\nu-1}(y_-)e^{ikL/2},
\eeax
so that
\be
\psi(z)=y_-\left[K_{\nu-1}(y_-)I_\nu(y)+I_{\nu-1}(y_-)K_\nu(y)\right]e^{ikL/2}
\ee
Again, the initial conditions are satisfied by virtue of Eq.~(\ref{Wronskian}).

At $z=L/2$ we have to match $\psi$ with $A_R e^{-ikL/2} +B_R e^{ikL/2}$.
The general formulas are
\beax
A_R e^{-ikz}&=& \half[\psi(z)+(i/k)\psi'(z)]\cr
&&\cr
B_R e^{ikz}&=& \half[\psi(z)-(i/k)\psi'(z)]
\eeax
giving
\bea\label{RT}
A_R&=& (\half \alpha^2 \nu) e^{ikL}[\phantom{-}I_{\nu-1}(y_-)K_{\nu+1}(y_+)-K_{\nu-1}(y_-)I_{\nu+1}(y_+)]\nonumber\\
&&\\
B_R&=& (\half \alpha^2 \nu) e^{-ikL}[-I_{\nu-1}(y_-)K_{\nu-1}(y_+)+K_{\nu-1}(y_-)I_{\nu-1}(y_+)]\nonumber
\eea
Note that the reflection amplitude $B_R$ can be obtained from $B_L$ by the transformations $y_+ \leftrightarrow y_-$ and $\nu \leftrightarrow -\nu$, which is a consequence of $PT$ symmetry. Under those same transformations the transmission amplitude $A_L\equiv A_R$ is invariant. The equality of $A_L$ and $A_R$ is a general result, obtained most easily by evaluating the Wronskian of the two solutions $\psi_L(z)$ and $\psi_R(z)$ for $z < -L/2$ and $z>L/2$.

A caution about the implementation of Eqs.~(\ref{LT}) and (\ref{RT}) is in order. As noted in Ref.~\cite{HFJ1}, the argument $y$ of the Bessel functions encircles the origin and crosses the cut on the negative real axis many times as $z$ goes from $-L/2$ to $L/2$. Thus it is important to know how to continue onto subsequent sheets. Once the relevant formulas given in Ref.~\cite{HFJ1} are implemented the resulting functions are smooth functions of $z$, with no discontinuities.
\section{Analytic Solution in General Case}\label{New}
In this section we generalize the previous results in two ways, by considering non-normal incidence
allowing for different background refractive indices on either side of the grating.

\subsection{Left Incidence}
The set-up for left incidence is shown in Fig.~3.
\begin{center}
\begin{figure}[h!]
\resizebox{!}{1.7in}{\includegraphics{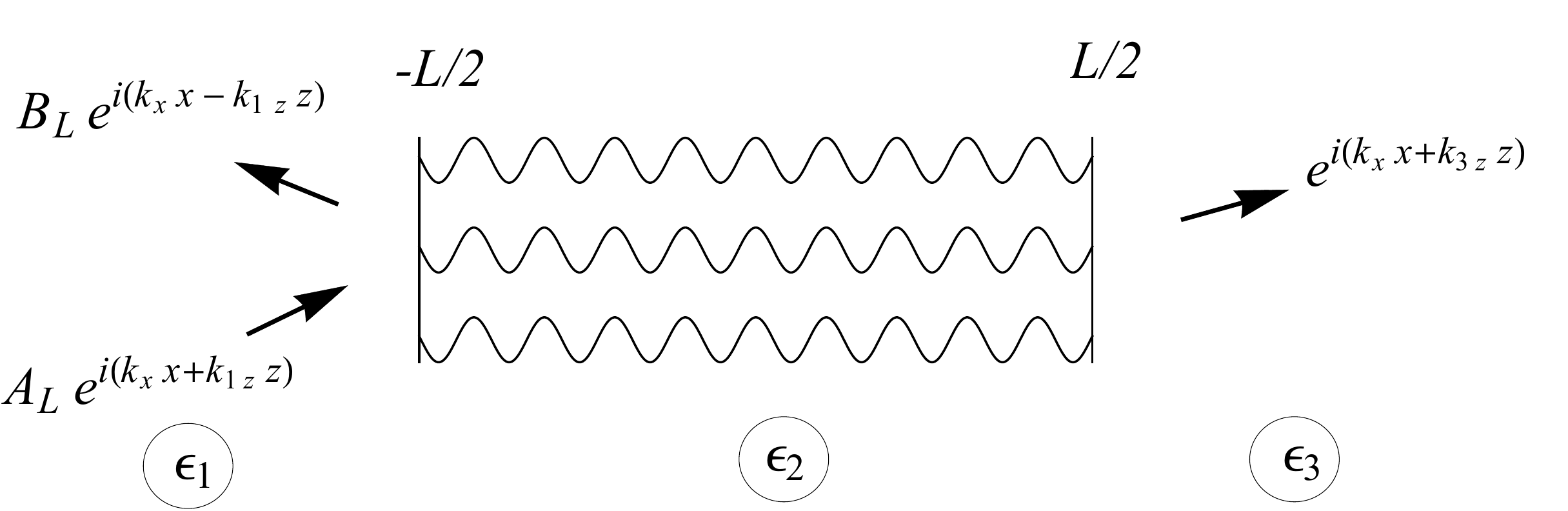}}
\caption{Generalized set-up for propagation from the left}
\end{figure}
\end{center}
For incidence at an angle, there is an overall component of the wave, in, say, the $x$-direction, of the form $e^{ik_x x}$. So in Eq.~(\ref{Helmholtz}) we can write $E$ as $E(x,z)=e^{ik_x x}\psi(z)$, and the equation for $\psi(z)$ becomes
\be
\frac{d^2\psi}{dz^2}+\left[k_2^2(1+\alpha^2 e^{2i\beta z})-k_x^2\right]\psi=0.
\ee
Here $k_2=\sqrt{\ve_2}k_0$, where $k_0=2\pi/\lambda$ is the free-space wave-vector. The appropriate definition of $y$ is now
$y=(k_2\alpha/\beta)e^{i\beta z}$, which again results in the modified Bessel equation
\be
y^2\frac{d^2\psi}{dy^2}+y\frac{d\psi}{dy}-(y^2+\nu^2)\psi=0,
\ee
with the difference that $\nu$ is now defined by
\bea
\nu^2=\frac{k_2^2-k_x^2}{\beta^2}
\eea
or in other words $\nu=k_{2z}/\beta=(k_2 \cos\theta)/\beta$, where $\theta$ is the internal angle of refraction. Thus $y$ can be written as
\be
y=\left(\frac{\nu\alpha}{\cos\theta}\right)e^{i\beta z}.
\ee
When considering non-normal incidence the boundary conditions depend on the polarization state of the input radiation. In this paper we will restrict ourselves to the simplest case of $H$-mode polarization, in which the $\boldsymbol{E}$-field is in the $y$-direction, and so parallel to the two interfaces at $z=\pm L/2$. In general the boundary conditions require that the tangential electric and magnetic fields be continuous across a boundary. In this case $H_x$ is proportional to $\psi'$, so the conditions are that $\psi$ and $\psi'$ be continuous, as before. The difference is that, for $\ve_2\ne \ve_1$ and/or $\ve_2\ne \ve_3$ the longitudinal wave-vectors in the three regions are unequal. Thus,
given that $k_1\sin\theta'=k_2\sin\theta$, it is straightforward to show that $k_{1z}=\gamma_1\nu\beta$ and $k_{3z}=\gamma_3\nu\beta$,
where $\gamma_r=\left(\ve_r/\ve_2-\sin^2\theta\right)^\half/\cos\theta$. In the previous case of equal background permittivities these
reduced to $\gamma_1=\gamma_2=1$.

Applying the boundary conditions of continuity for both $\psi$ and $\psi'$ we obtain, after some algebra, the following expressions for $A_L$ and $B_L$:
\bea\label{exactL}
A_L e^{-ikL}&=& \phantom{+}\left(\frac{y_+ y_-}{2 \gamma_1 \nu}\right)[I_{\nu-1}(y_-)K_{\nu+1}(y_+)-I_{\nu+1}(y_+)K_{\nu-1}(y_-)]\cr
&&+ \left(\frac{\delta_1 y_+}{2 \gamma_1}\right)[I_{\nu}(y_-)K_{\nu+1}(y_+)+I_{\nu+1}(y_+)K_{\nu}(y_-)]\cr
&&+ \left(\frac{\delta_3 y_-}{2 \gamma_1 }\right)[I_{\nu-1}(y_-)K_{\nu}(y_+)+I_{\nu}(y_+)K_{\nu-1}(y_-)]\cr
&&+  \left(\frac{\delta_1 \delta_3 \nu}{2 \gamma_1}\right)[I_{\nu}(y_-)K_{\nu}(y_+)-I_{\nu}(y_+)K_{\nu}(y_-)]\cr
&&\\
B_L e^{ikL}&=& -\left(\frac{y_+ y_-}{2 \gamma_1 \nu}\right)[I_{\nu+1}(y_-)K_{\nu+1}(y_+)-I_{\nu+1}(y_+)K_{\nu+1}(y_-)]\cr
&&+ \left(\frac{\delta_1 y_+}{2 \gamma_1}\right)[I_{\nu}(y_-)K_{\nu+1}(y_+)+I_{\nu+1}(y_+)K_{\nu}(y_-)]\cr
&&- \left(\frac{\delta_3 y_-}{2 \gamma_1 }\right)[I_{\nu+1}(y_-)K_{\nu}(y_+)+I_{\nu}(y_+)K_{\nu+1}(y_-)]\cr
&&+  \left(\frac{\delta_1 \delta_3 \nu}{2 \gamma_1}\right)[I_{\nu}(y_-)K_{\nu}(y_+)-I_{\nu}(y_+)K_{\nu}(y_-)]\nonumber
\eea
where $\delta_r=\gamma_r-1$. Each expression now has potentially three additional terms due to the fact that in the general case $\delta_r\ne 0$.

\subsection{Right Incidence}
The set-up for right incidence is shown in Fig.~4.
\begin{center}
\begin{figure}[h!]
\resizebox{!}{1.7in}{\includegraphics{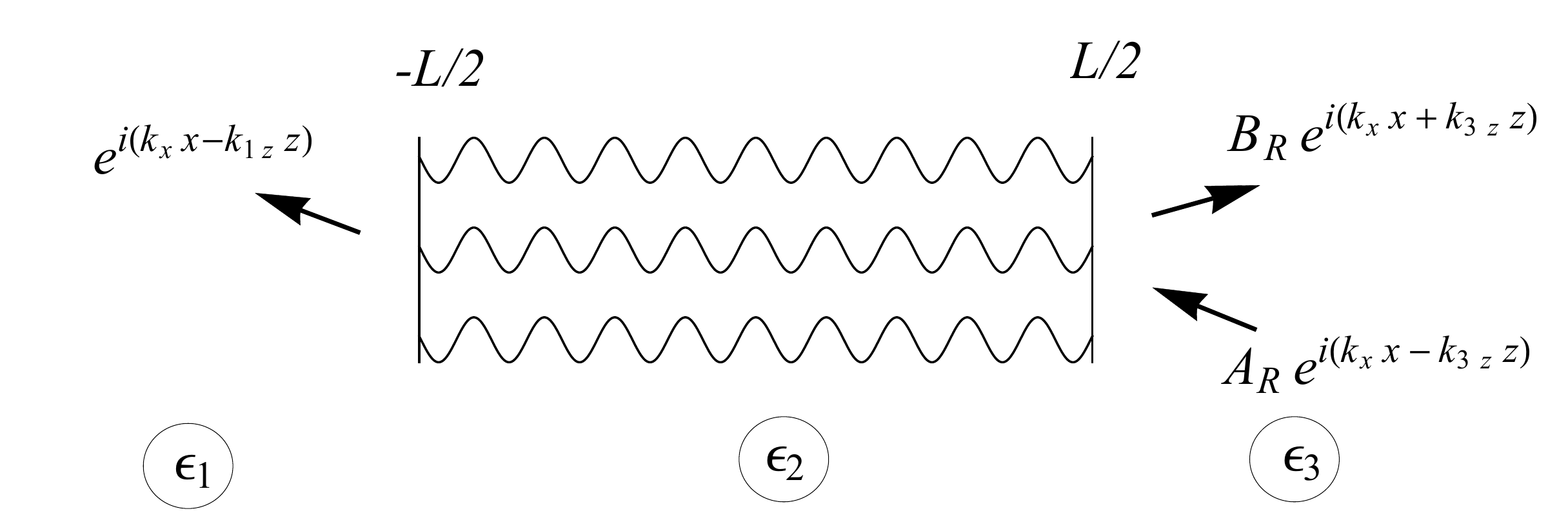}}
\caption{Generalized set-up for propagation from the right}
\end{figure}
\end{center}
The algebra follows on similar lines and results in the following expression for $A_R$, $B_R$:
\bea\label{exactR}
A_R e^{ikL}&=& \phantom{+}\left(\frac{y_+ y_-}{2 \gamma_3 \nu}\right)[I_{\nu-1}(y_-)K_{\nu+1}(y_+)-I_{\nu+1}(y_+)K_{\nu-1}(y_-)]\cr
&&+ \left(\frac{\delta_1 y_+}{2 \gamma_3}\right)[I_{\nu}(y_-)K_{\nu+1}(y_+)+I_{\nu+1}(y_+)K_{\nu}(y_-)]\cr
&&+ \left(\frac{\delta_3 y_-}{2 \gamma_3 }\right)[I_{\nu-1}(y_-)K_{\nu}(y_+)+I_{\nu}(y_+)K_{\nu-1}(y_-)]\cr
&&+  \left(\frac{\delta_1 \delta_3 \nu}{2 \gamma_3}\right)[I_{\nu}(y_-)K_{\nu}(y_+)-I_{\nu}(y_+)K_{\nu}(y_-)]\cr
&&\\
B_R e^{ikL}&=& \phantom{+}\left(\frac{y_+ y_-}{2 \gamma_3 \nu}\right)[I_{\nu-1}(y_-)K_{\nu+1}(y_+)-I_{\nu+1}(y_+)K_{\nu-1}(y_-)]\cr
&&+ \left(\frac{\delta_1 y_+}{2 \gamma_3}\right)[I_{\nu}(y_-)K_{\nu+1}(y_+)+I_{\nu+1}(y_+)K_{\nu}(y_-)]\cr
&&- \left(\frac{\delta_3 y_-}{2 \gamma_3 }\right)[I_{\nu-1}(y_-)K_{\nu}(y_+)+I_{\nu}(y_+)K_{\nu-1}(y_-)]\cr
&&+  \left(\frac{\delta_1 \delta_3 \nu}{2 \gamma_3}\right)[I_{\nu}(y_-)K_{\nu}(y_+)-I_{\nu}(y_+)K_{\nu}(y_-)]\nonumber
\eea
Note that, with different background permittivities $\ve_1$ and $\ve_3$, the relation between $A_R$ and $A_L$ is now $|A_L|=(\gamma_3/\gamma_1)|A_R|=(k_{3z}/k_{1z})|A_R|$. This relation can again be obtained by considering the Wronskian of $\psi_L(z)$ and $\psi_R(z)$.
\section{Numerical Results}\label{Numerical}
In this section we explore a variety of different configurations. It should be stressed again that these are configurations which have  previously been considered within the Bragg approximation scheme in Ref.~\cite{RG}. The results can be directly compared\footnote{However, the reader should be aware that the convention for a plane wave used in that paper, $e^{-j k z}$, is opposite to the one used here, namely $e^{i k z}$, so that left and right are effectively interchanged.}, and are broadly similar but differ in some details. This is to expected, given the relatively small strength ($\alpha^2$) of the grating. For larger grating strengths only the present method can be expected to give reliable results, unless several more orders in the Bragg series can be included.
\subsection{Filled-space grating}
As the first application of the equations we have derived, we consider oblique incidence on the grating, but keeping the background relative permittivities the same, as in Ref.~\cite{HFJ1}. In Fig.~5 (left panel) we show the transmission coefficient, which, as already remarked, is the same for left or right incidence. The same characteristic shape that was seen in Ref.~\cite{HFJ1} is seen again here, but this time as a function of $\theta$ rather than $k$. The transition occurs near $\theta_B = \arccos(\beta/(k\sqrt{\ve_2}))=\arccos(\lambda/(2\Lambda\sqrt{\ve_2}))\approx 1.06$, the angle for which $\nu=1$. In the right panel we show the left reflection coefficient $R_L$, which is small, although it increases with larger $|\theta|$, and shows no transition near $\theta=\pm\theta_B$. Thus invisibility from the left is preserved to a large extent, with a small reflection coefficient and a transmission coefficient very close to 1.
\begin{center}
\begin{figure}[h!]
\resizebox{!}{1.7in}{\includegraphics{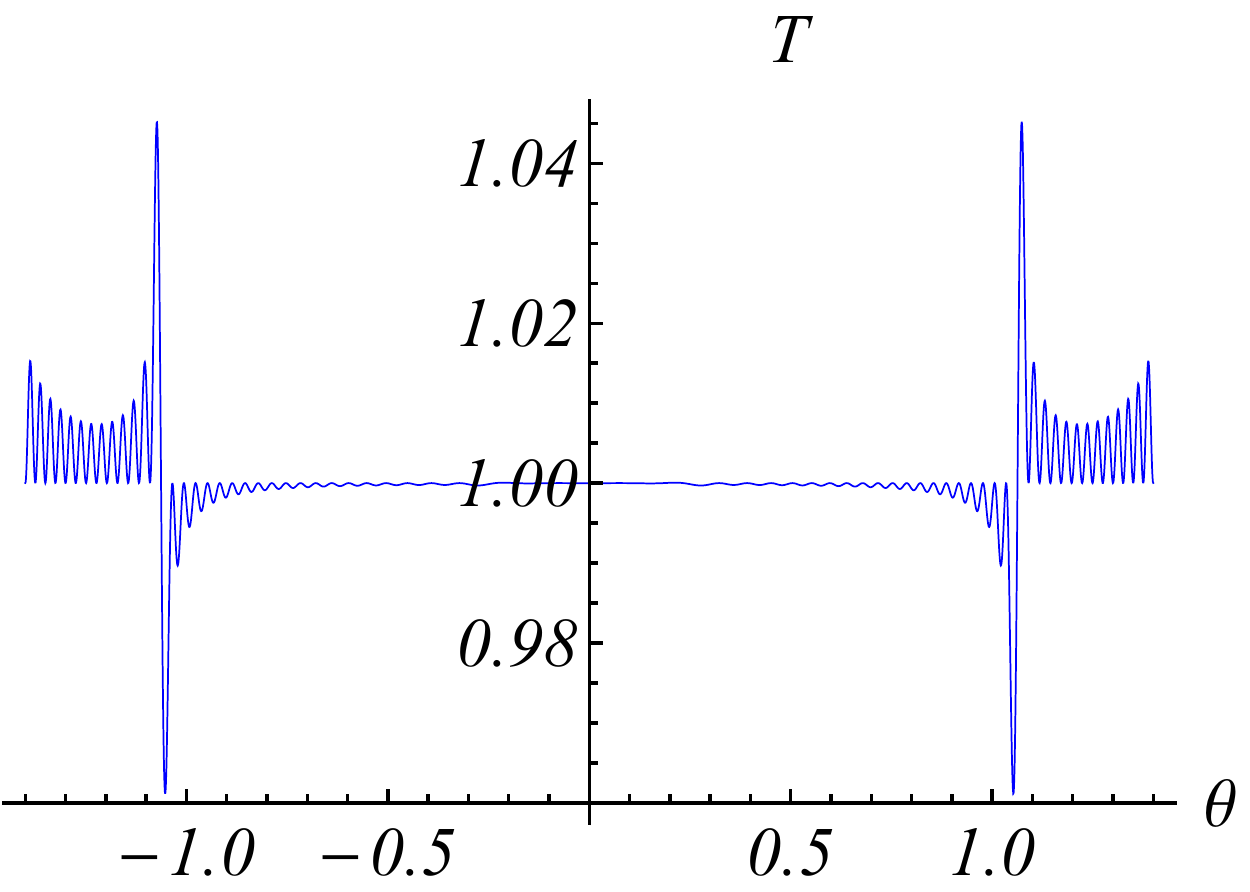}}\hspace{0.5in}\resizebox{!}{2in}{\includegraphics{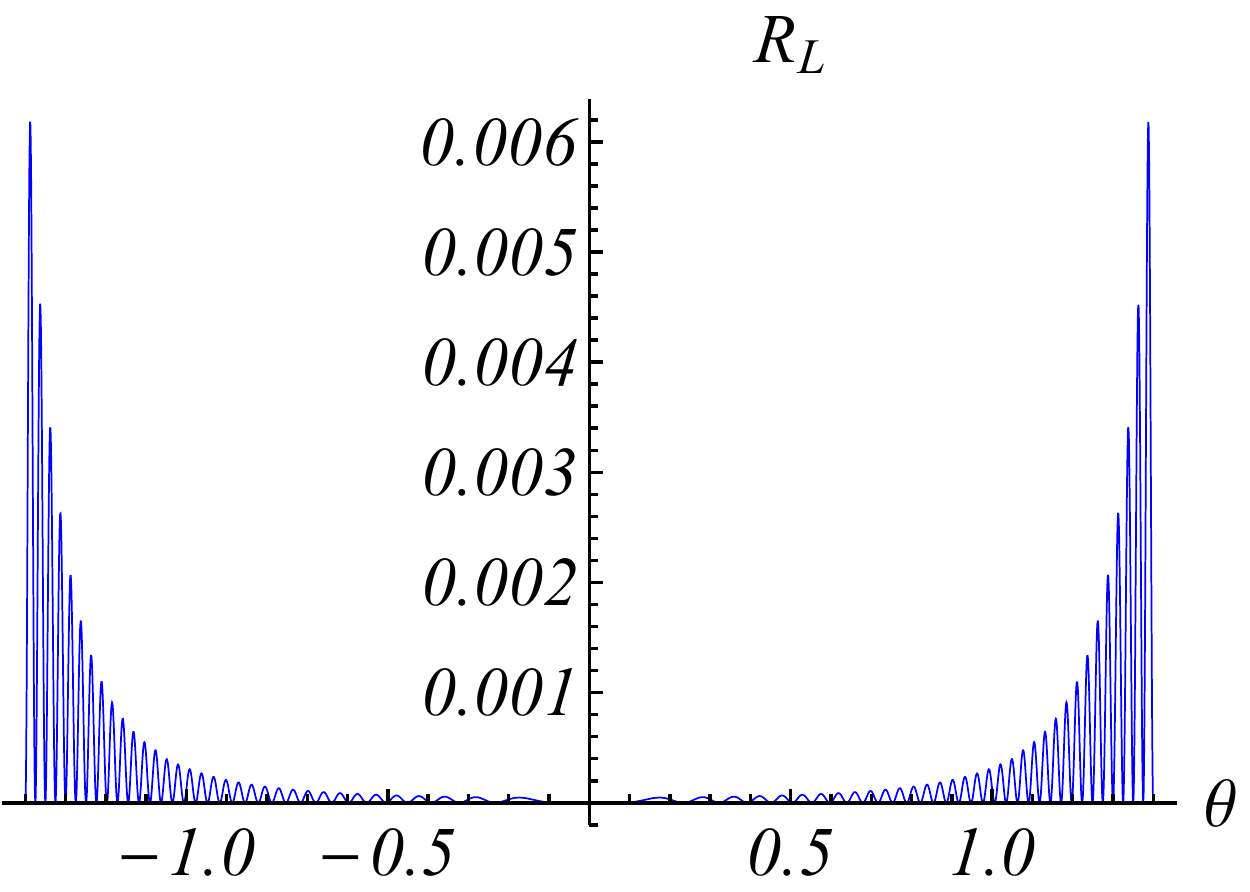}}
\caption{Transmission coefficient $T_L$ ($=T_R$) and reflection coefficient $R_L$ as functions of the internal angle of refraction $\theta$ for the case
$\ve_1=\ve_2=\ve_3=2.4$. The other parameters are $\alpha^2=0.02$,
$L = 8.4$, $\Lambda =0.42$ and $\lambda=0.633$.}
\end{figure}
\end{center}
In Fig.~6 we show the right reflection coefficient $R_R$ for the same set of parameters.  As a function of $\theta$ the right reflection coefficient displays the same high narrow peak that occurs for normal incidence when $k$ is varied. It should be mentioned that these reflection and transmission coefficients satisfy the modified unitarity relation $T-1=\pm \sqrt{R_L R_R}$ of Ref.~\cite{Ge} to a high degree of accuracy, which provides a stringent test of our formulas. Note that $T\gtrless 1$ for $|\theta|\gtrless|\theta_B|$.
\begin{center}
\begin{figure}[h!]
\resizebox{!}{1.7in}{\includegraphics{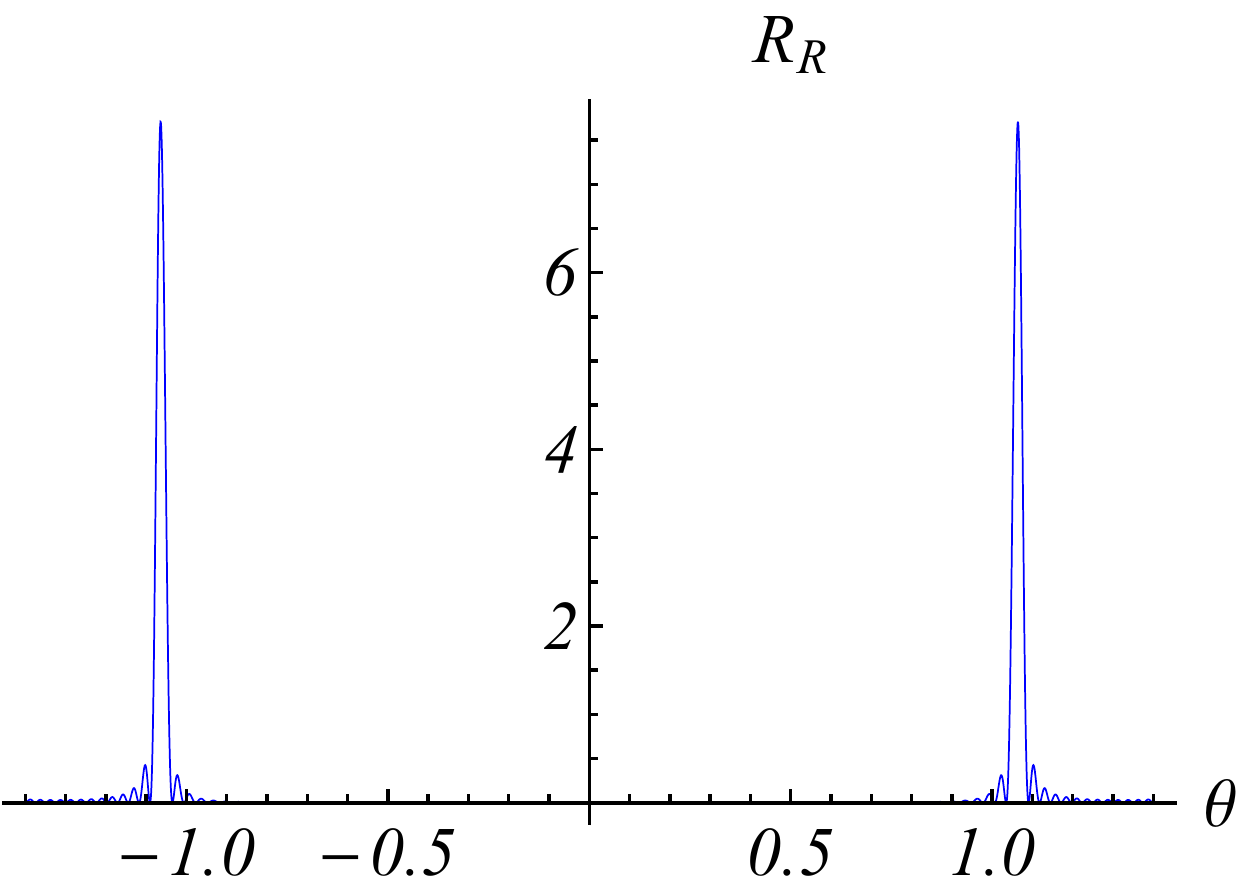}}
\caption{Right reflection coefficient as a function of the internal angle of refraction $\theta$ for the parameters of Fig.~5}
\end{figure}
\end{center}
\subsection{Grating on a slab in air}
A rather natural set-up would be for the optical lattice/grating to be implemented on a slab of material with background permittivity $\ve_2$ different from the permittivities on either side. As an example, in this subsection we consider a slab in air, with $\ve_1=\ve_3=1$. It is only to be expected that the reflection and transmission properties will be significantly modified in this case, due to reflections at the interfaces between the different materials. This is indeed borne out by Fig.~7, for the transmission coefficient and the left reflection coefficient.
\begin{center}
\begin{figure}[h!]
\resizebox{!}{1.7in}{\includegraphics{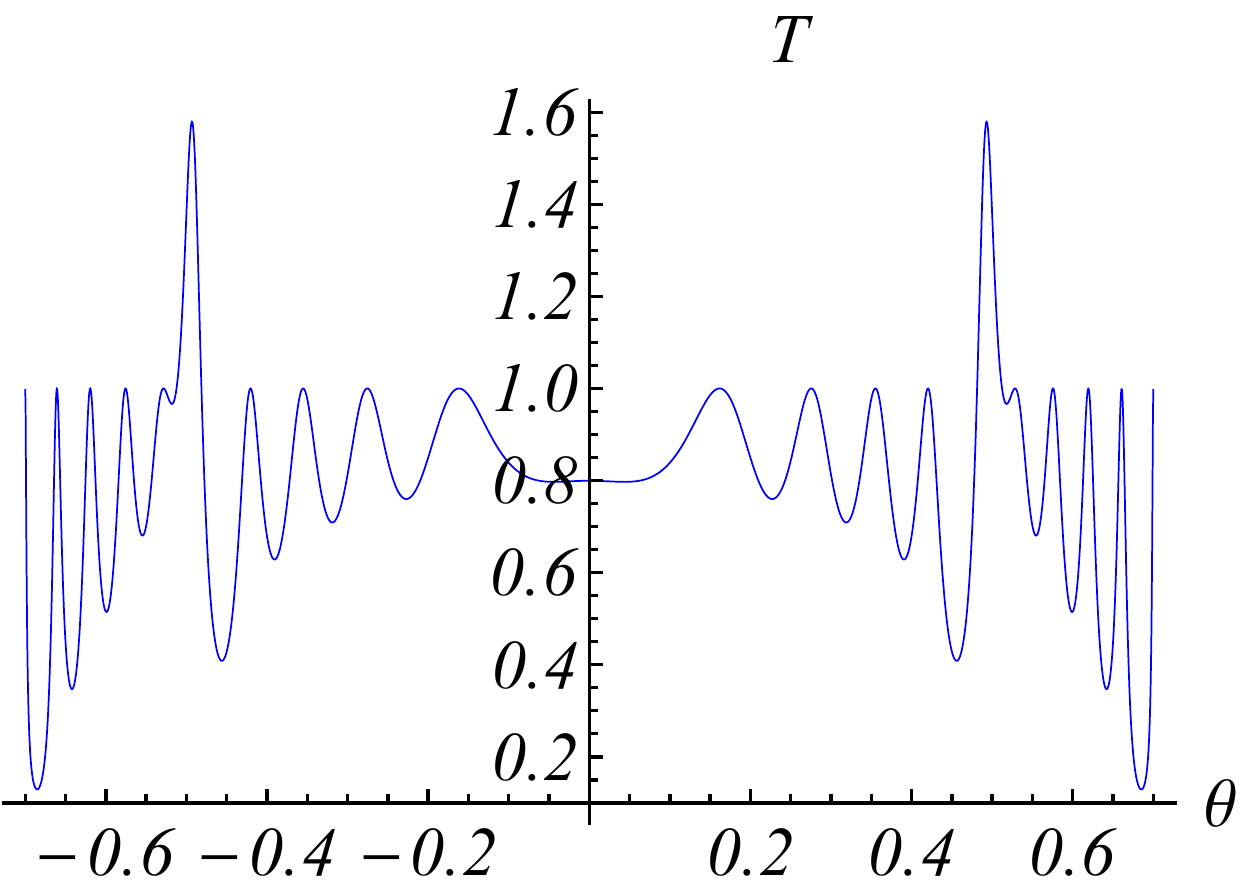}}\hspace{0.5in}\resizebox{!}{1.7in}{\includegraphics{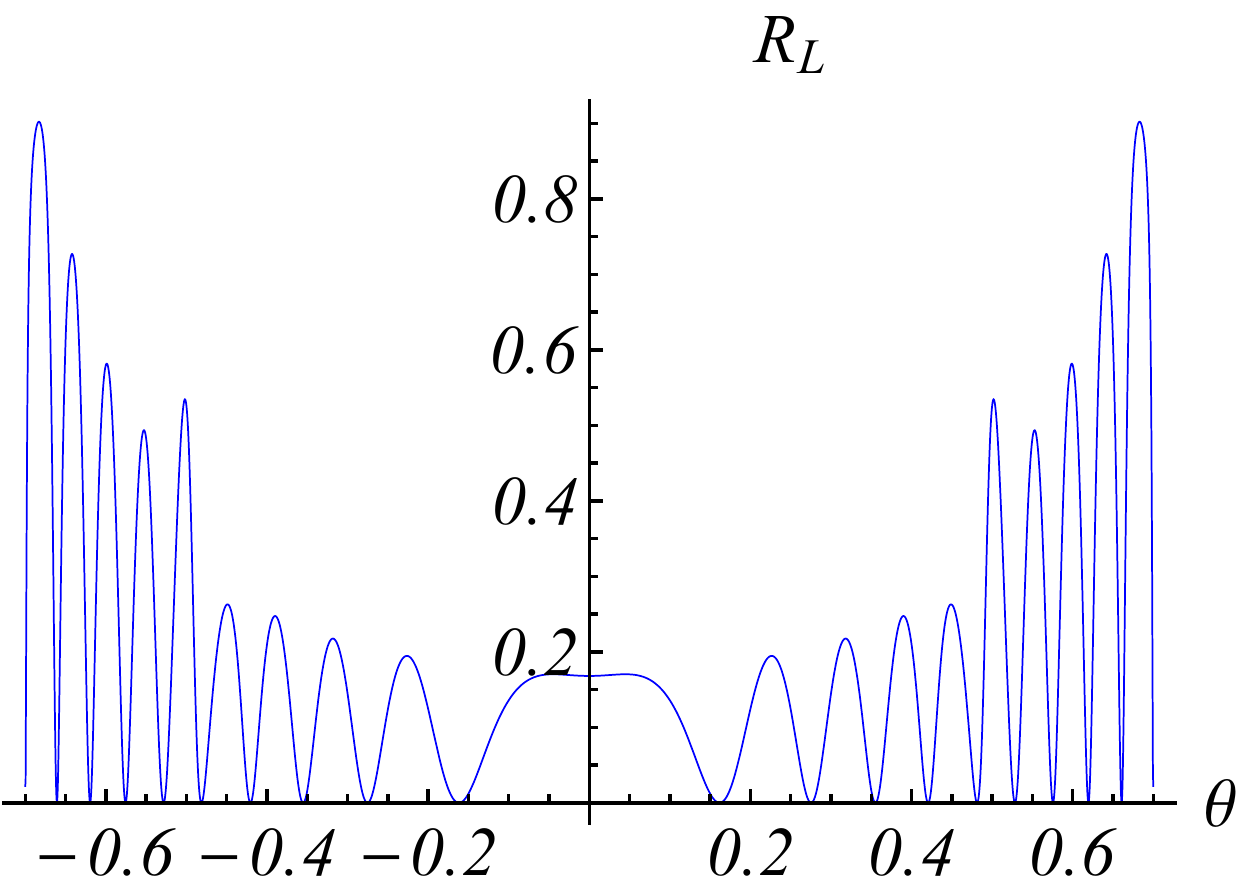}}
\caption{Transmission coefficient $T_L$ ($=T_R$) and reflection coefficient $R_L$ as functions of the internal angle of refraction $\theta$ for the case $\ve_1=\ve_3=1$, $\ve_2=2.4$. The other parameters are $\alpha^2=0.02$,
$L = 8.28$, $\Lambda =0.23$ and $\lambda=0.633$.}
\end{figure}
\end{center}
However, a more relevant quantity as far as unidirectional invisibility is concerned is the contrast, i.e. the differences $\Delta T \equiv T-T_0$ and $\Delta R_L \equiv R_L-R_0$ where $T_0$ and $R_0$ are the transmission coefficients in the absence of the grating, i.e. with $\alpha=0$. These quantities are shown in Fig.~8, which reveals that $\Delta T$ is rather small for $|\theta|\lesssim 0.3$, but becomes appreciable for larger values, and indeed becomes of $O(1)$ in the vicinity of $|\theta| = \theta_B$. The contrast $\Delta R_L$ is quite small overall, particularly for $|\theta|\lesssim 0.3$, but shows a significant peak in the vicinity of $|\theta| = \theta_B$.
\begin{center}
\begin{figure}[h!]
\resizebox{!}{1.7in}{\includegraphics{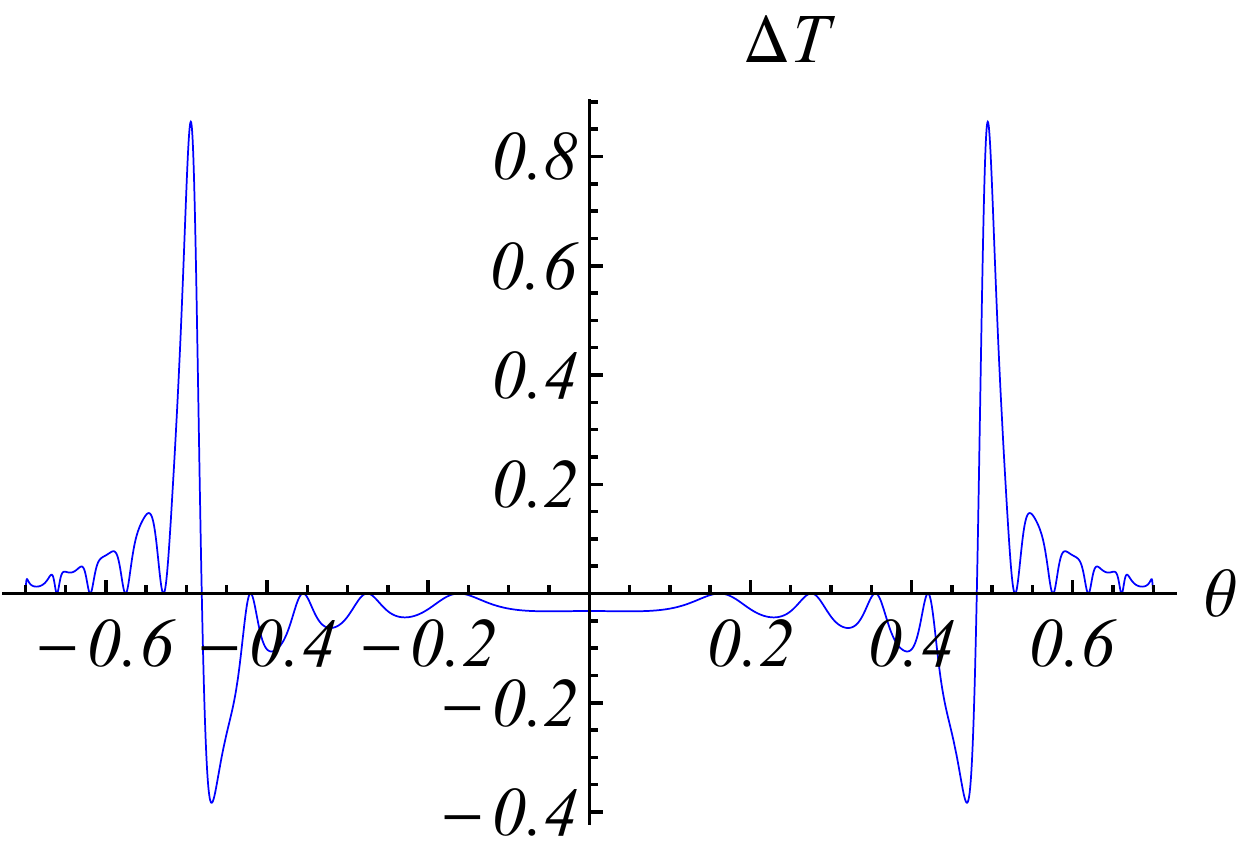}}\hspace{0.5in}\resizebox{!}{1.7in}{\includegraphics{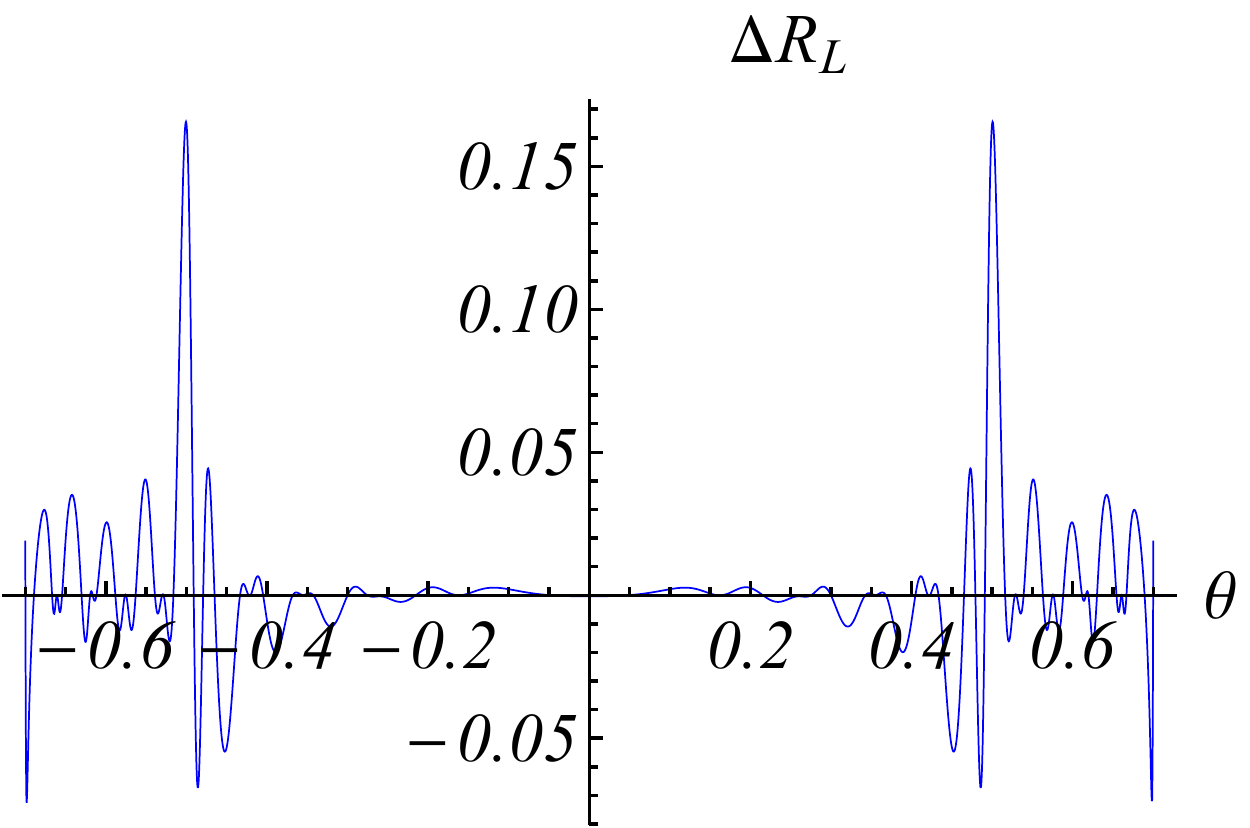}}
\caption{Contrasts $\Delta T_L$ ($=\Delta T_R$) and $\Delta R_L$ as functions of the internal angle of refraction $\theta$ for the parameters of Fig.~7.}
\end{figure}
\end{center}
In Fig.~9 we show the right reflection coefficient, which still shows a characteristic peak near $|\theta| = \theta_B$, but other structure arising from reflection from the boundaries besides.
\begin{center}
\begin{figure}[h!]
\resizebox{!}{2in}{\includegraphics{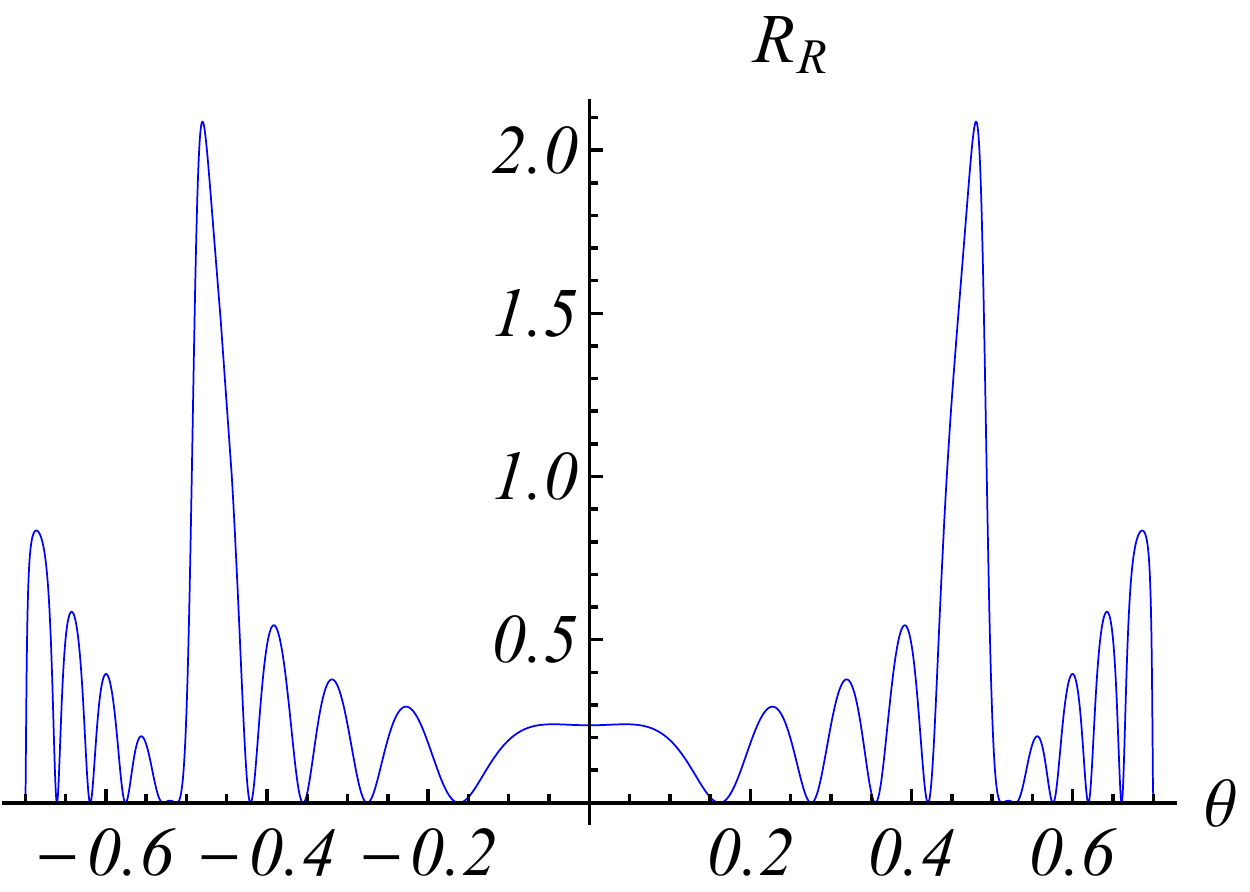}}
\caption{Right reflection coefficient $R_R$ as a function of the internal angle of refraction $\theta$ for the parameters of Fig.~7.}
\end{figure}
\end{center}
\subsection{Grating on a substrate}
Another natural situation in practice could be for the grating to be implemented on a substrate with a different background permittivity. So, for example, we could have $\ve_1=1$, $\ve_2=2.4$ and $\ve_3=2$, in which case the reflective side of the grating would be attached
to the substrate. We could also have the non-reflective side of the grating attached to the substrate, a situation that was considered in Ref.~\cite{RG}, but which we will not discuss here. In this asymmetric case, when $\ve_1 \ne \ve_3$, the set-up is no longer $PT$ symmetric, so that the test of generalized unitarity cannot be applied, and also the relation $T_L=T_R$ no longer holds, as has been previously remarked. One therefore defines the diffraction efficiency for transmission from the left as $\hat{T}_L=(k_{3z}/k_{1z})T_L$ and similarly $\hat{T}_R=(k_{1z}/k_{3z})T_L$, so that $\hat{T}_L=\hat{T}_R$.

Again the reflection and transmission properties are modified, although rather less than in Case B. In Fig.~10 we show the diffaction efficiency for transmission and the left reflection coefficient.
\begin{center}
\begin{figure}[h!]
\resizebox{!}{2in}{\includegraphics{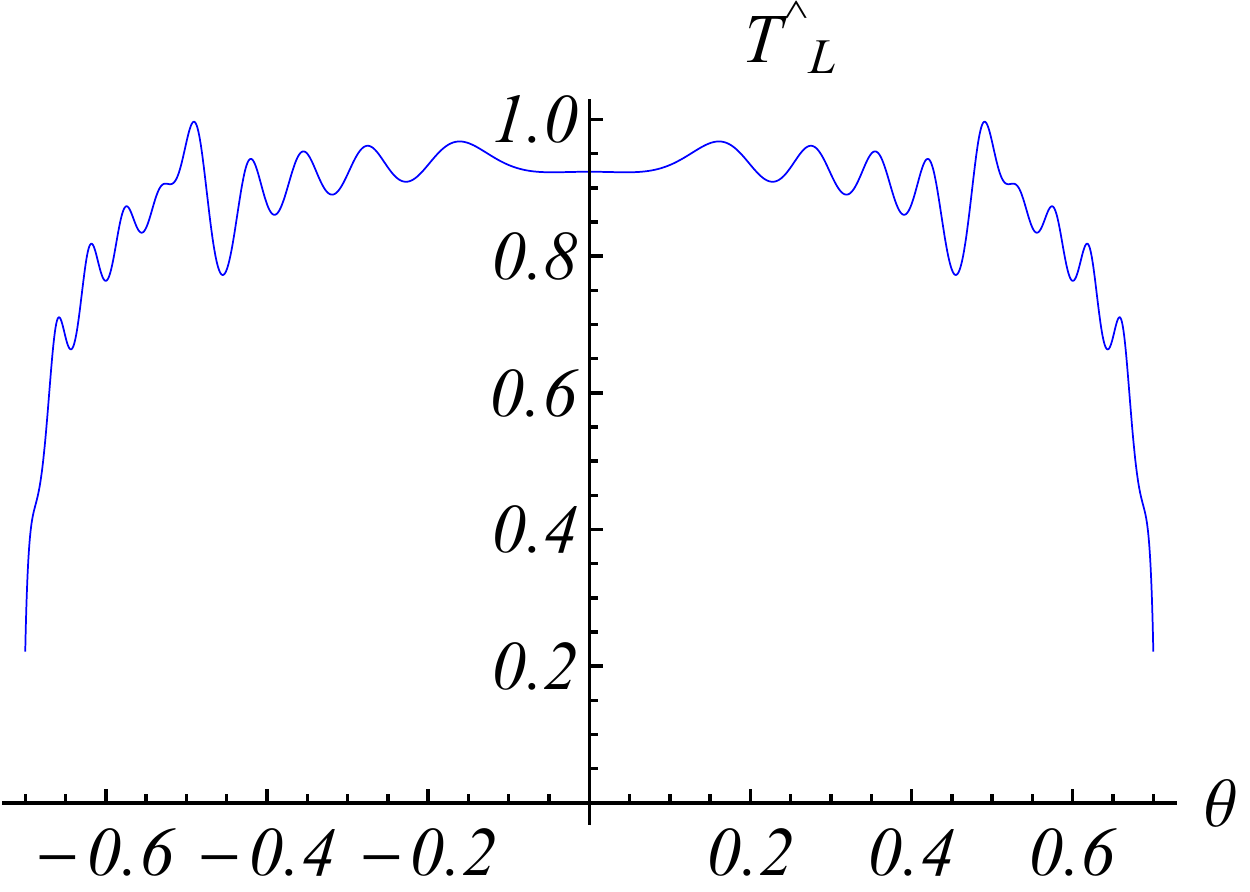}}\hspace{0.5in}\resizebox{!}{2in}{\includegraphics{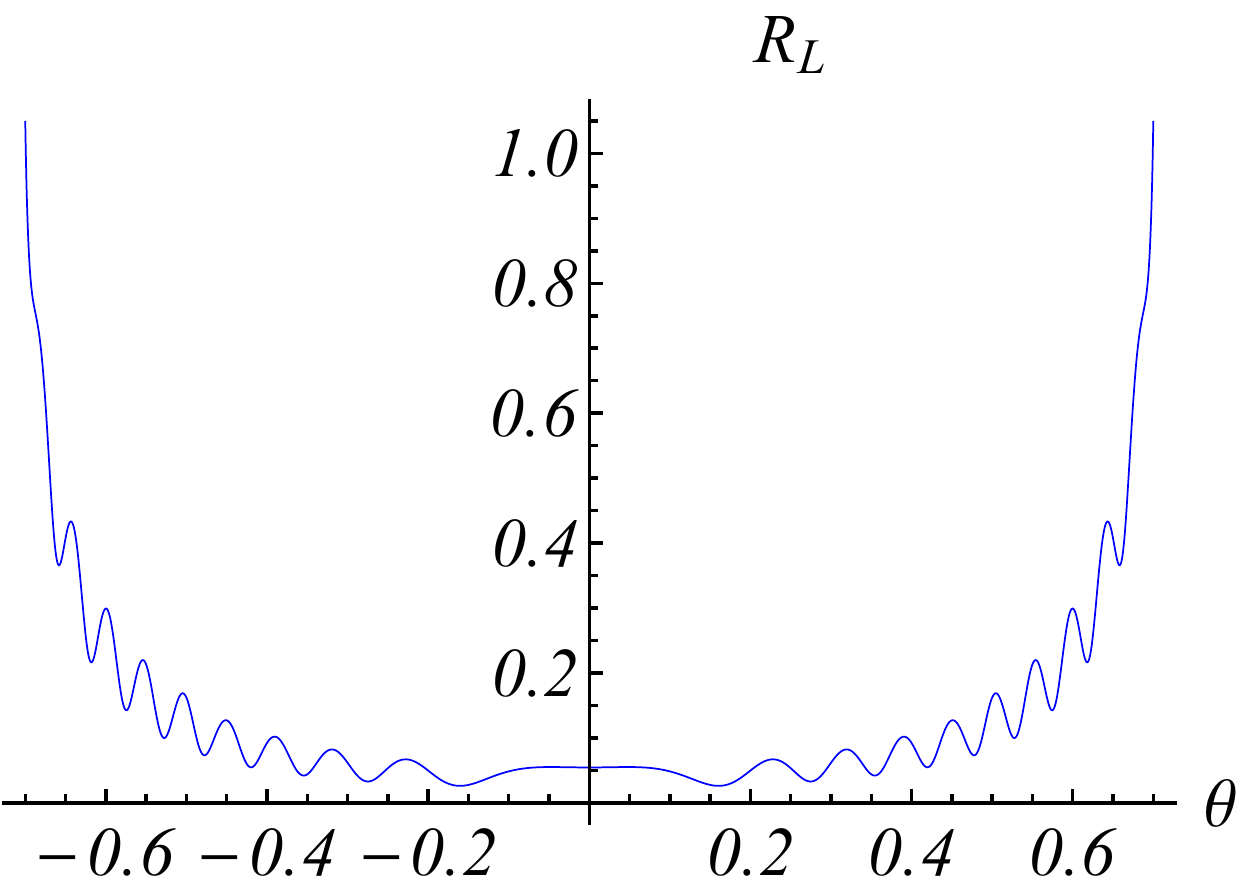}}
\caption{Diffraction efficiency for transmission $\hat{T}_L$ ($=\hat{T}_R$) and reflection coefficient $R_L$ as functions of the internal angle of refraction $\theta$ for the case $\ve_1=1$, $\ve_2=2.4$, $\ve_3=2$,. The other parameters are as in Fig.~7.}
\end{figure}
\end{center}
The corresponding contrasts
 $\Delta\hat{T} \equiv \hat{T}-\hat{T}_0$ and $\Delta R_L \equiv R_L-R_0$  are shown in Fig.~11, from which it can be seen that $|\Delta \hat{T}|\lesssim 0.1$, and is considerably smaller in the region $|\theta|\lesssim 0.3$. The contrast $\Delta R_L$ is an order of magnitude smaller.
\begin{center}
\begin{figure}[h!]
\resizebox{!}{1.9in}{\includegraphics{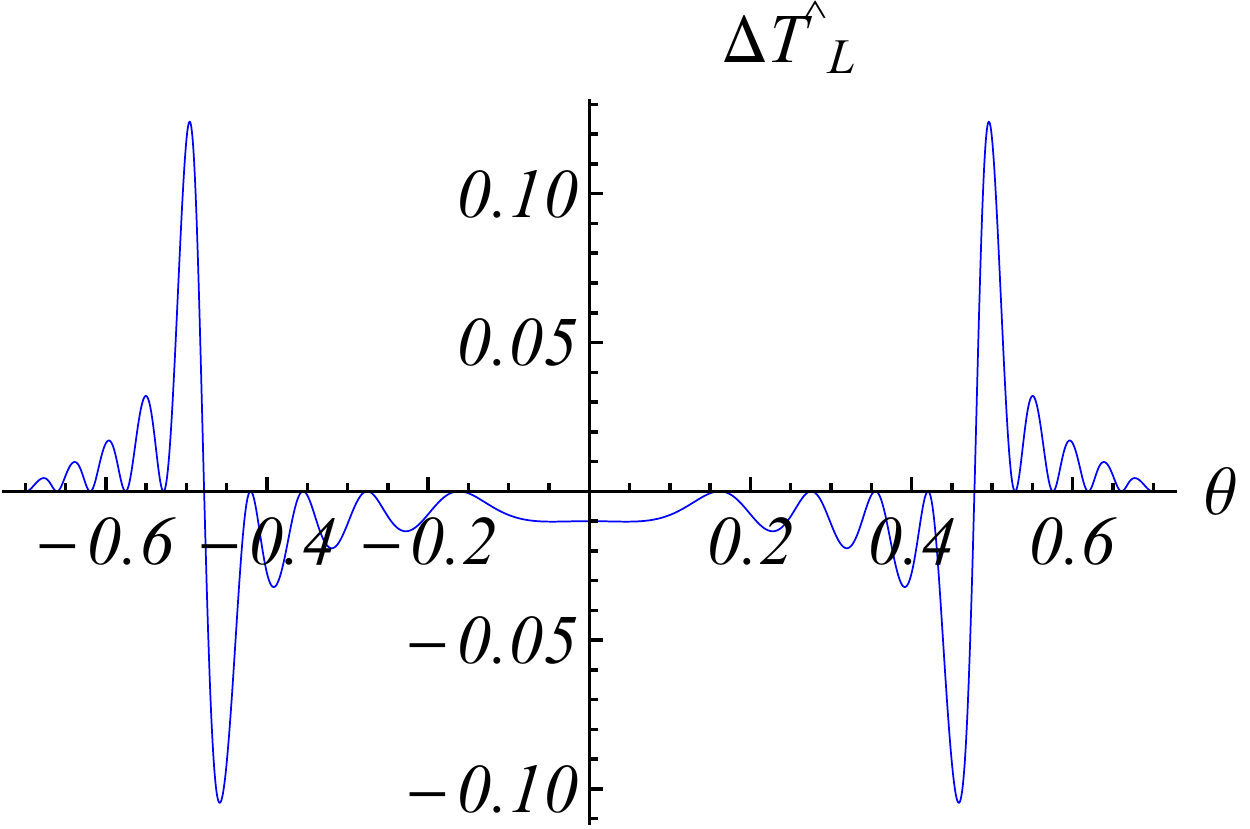}}\hspace{0.5in}\resizebox{!}{1.9in}{\includegraphics{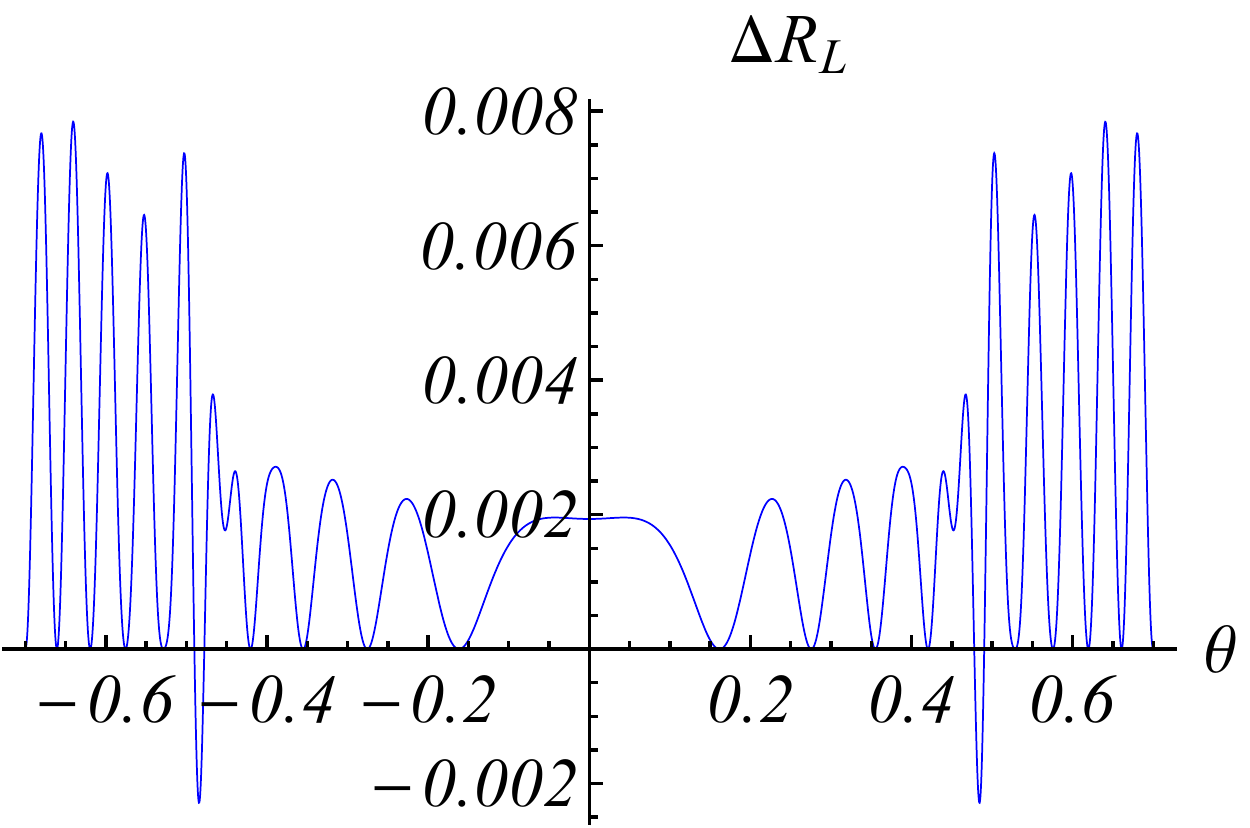}}
\caption{Contrasts $\Delta \hat{T}_L$ ($=\Delta \hat{T}_R$) and $\Delta R_L$ as functions of the internal angle of refraction $\theta$ for the parameters of Fig.~10.}
\end{figure}
\end{center}
In Fig.~12 we show the right reflection coefficient, which now has a fairly clean peak near $|\theta| = \theta_B$, with some additional structure for larger $|\theta|$.
\begin{center}
\begin{figure}[h!]
\resizebox{!}{2.5in}{\includegraphics{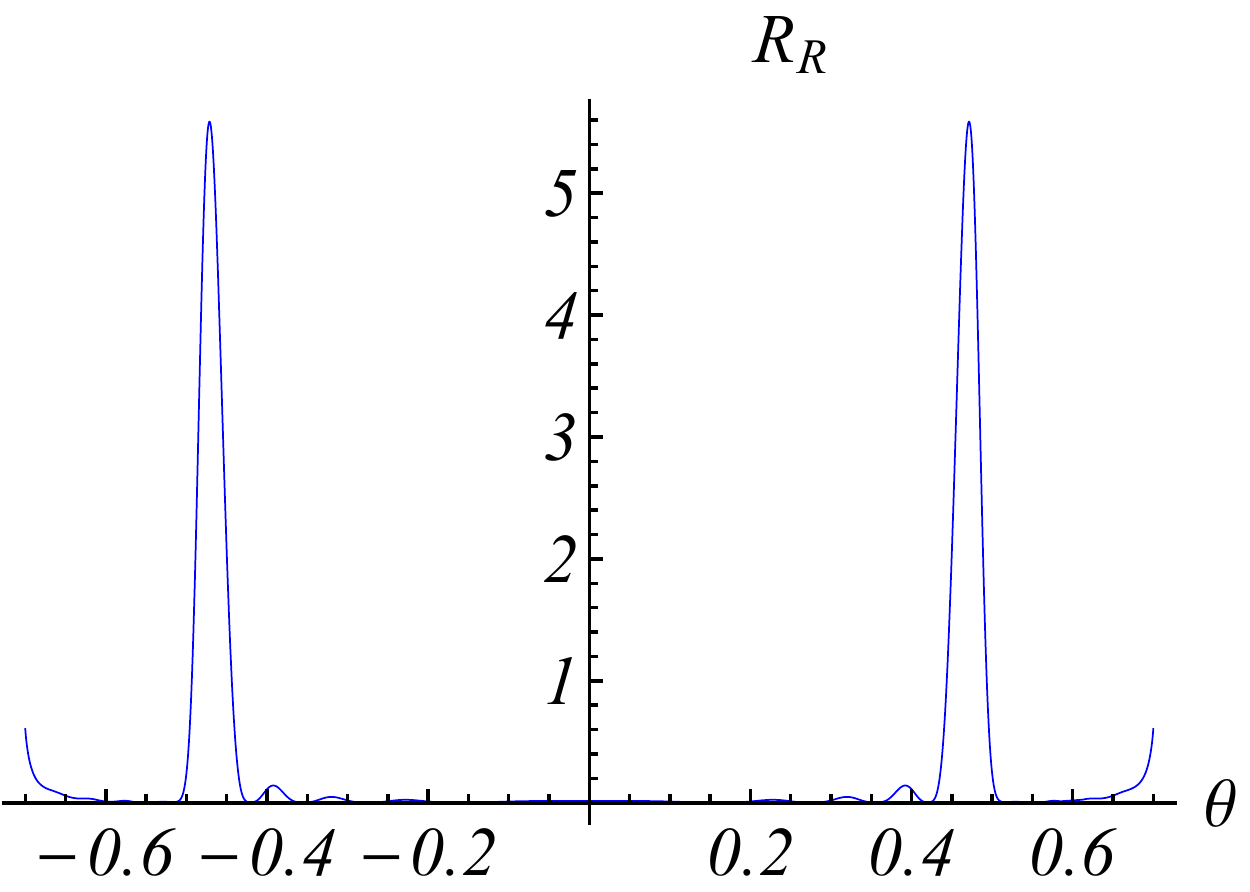}}
\caption{Right reflection coefficient $R_R$ as a function of the internal angle of refraction $\theta$ for the parameters of Fig.~10.}
\end{figure}
\end{center}
\section{Mirror Set-up}\label{Mirror}
In this case we revert to normal incidence from the left and take $\ve_3 \to \infty$. Then both $A_L$ and $B_L$ tend to infinity because of the terms containing the factor $\delta_3$. We can take out this factor and define $\hat{A}_L = A_L/\delta_3$ and $\hat{B}_L = B_L/\delta_3$, which in this limit are given by
\bea
\hat{A}_L e^{ikL}&=&\left(\frac{ y_-}{2 \gamma_1 }\right)[I_{\nu-1}(y_-)K_{\nu}(y_+)+I_{\nu}(y_+)K_{\nu-1}(y_-)]\cr
&&+  \left(\frac{\delta_1 \nu}{2 \gamma_1}\right)[I_{\nu}(y_-)K_{\nu}(y_+)-I_{\nu}(y_+)K_{\nu}(y_-)]
\eea
 and
\bea
\hat{B}_L e^{ikL}&=& - \left(\frac{ y_-}{2 \gamma_1 }\right)[I_{\nu+1}(y_-)K_{\nu}(y_+)+I_{\nu}(y_+)K_{\nu+1}(y_-)]\cr
&&+  \left(\frac{\delta_1  \nu}{2 \gamma_1}\right)[I_{\nu}(y_-)K_{\nu}(y_+)-I_{\nu}(y_+)K_{\nu}(y_-)]
\eea

Because of the enhanced reflectivity of the grating for right incidence narrowly peaked around $k_2=\beta$, it seems likely that we will have a resonant cavity near the corresponding frequency, which might well support lasing. For lasing we are looking for a reflection coefficient $|B_L/A_L|=|\hat{B}_L/\hat{A}_L|$ going to infinity, so we are looking for a zero of $\hat{A}_L$, or equivalently of $A_R$. In Fig.~13 we show $R_L$ as a function of $k$ ($=2\pi/\lambda$) for the value of $\alpha$ we have been using throughout the paper. As we can see, this exhibits an extremely sharp peak at a certain value of $k$, indicating that for that value of $k$ we are very near a zero of $\hat{A}_L$. By fine-tuning $\alpha$ we can find a zero of $\hat{A}_L$ for real $k$. In fact, the complex zero in $k$ migrates from the lower half-plane to the upper half-plane as $\alpha$ increases, crossing the real axis at the critical value, the lasing threshold. We will be exploring this lasing set-up in more detail in future work\cite{HFJ&MK}.
\begin{center}
\begin{figure}[h!]
\resizebox{!}{2.5in}{\includegraphics{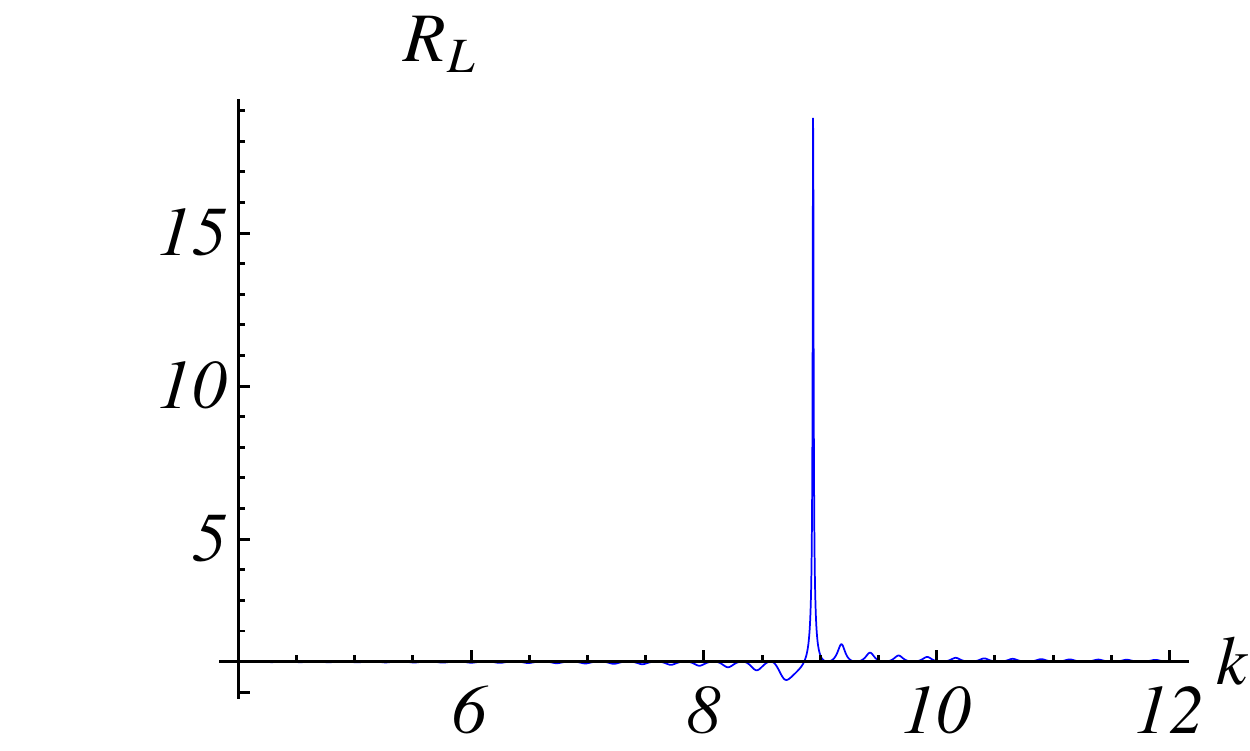}}
\caption{Left reflection coefficient $R_L$ for normal incidence as a function of $k$ for $\ve_1=1$, $\ve_2=2.4$ and $\ve_3 \to \infty$. The other parameters are $\alpha^2=0.02$, $L = 8.4$ and $\Lambda =0.42$.}
\end{figure}
\end{center}
\section{Conclusions}\label{Conclusions}
We have shown how the exact analytic solutions previously obtained for the one-dimensional grating with index profile proportional to $e^{i\beta z}$ can be generalized to the case of different refractive indices on either side of the grating as well as to non-normal incidence. The resulting formulas for the reflection and transmission amplitudes, Eqs.~(\ref{exactL}) and (\ref{exactR}), contain three additional terms, of the same general form as in the restricted case, but with different indices on the modified Bessel functions $I$ and $K$. These formulas were used to explore the transmission and reflection characteristics of the grating as a function of incident angle in a variety of situations previously considered within the Bragg approximation scheme, with particular emphasis on the extent to which unidirectional invisibility survives. The narrow-beam enhanced reflection of the grating for right incidence leads one to suppose that when a mirror is placed to the right of the cavity, the arrangement might support lasing, which calculation using our generalized formulas shows to be indeed the case.

\end{document}